\documentclass{article}
\usepackage{spconf,amsmath,epsfig}
\usepackage{multirow}
\newcommand{\atopp}[2]{\genfrac{}{}{0pt}{}{#1}{#2}}

\ninept
\title{Low-Rank STAP Algorithm for Airborne Radar Based on Basis-Function Approximation}
\name{Rui Fa, Rodrigo C. de Lamare and Sheng Li}
\address{\small Communications Research Group, Department of Electronics, \\
\small University of York, YO10 5DD, United Kingdom. \\ \small email: \{rf533, rcdl500\}@ohm.york.ac.uk, sl546@york.ac.uk}
\begin{document}
%
\maketitle
\begin{abstract}
In this paper, we develop a novel reduced-rank space-time adaptive processing (STAP) algorithm based on adaptive basis function approximation (ABFA) for airborne radar applications. The proposed algorithm employs the well-known framework of the side-lobe canceller (SLC) structure and consists of selected sets of basis functions that perform dimensionality reduction and an adaptive reduced-rank filter. Compared to traditional reduced-rank techniques, the proposed scheme works on an instantaneous basis, selecting the best suited set of basis functions at each instant to minimize the squared error. Furthermore, we derive stochastic gradient (SG) and recursive least squares (RLS) algorithm for efficiently implementing the proposed ABFA scheme. Simulations for a clutter-plus-jamming suppression application show that the proposed STAP algorithm outperforms the state-of-the-art reduced-rank schemes in convergence and tracking at significantly lower complexity.
\end{abstract}
\footnotetext{This work is funded by the Ministry of Defence (MoD), UK. Project MoD, Contract No. RT/COM/S/021.}
\section{Introduction}
\label{sec:intro}
Airborne surveillance radar systems operate in a severe and dynamic interference environments, which may be composed of clutter and jamming. Space-time adaptive processing (STAP) employing a joint-domain optimization of the spatial and temporal degrees-of-freedom (DOFs) has become a key enabling technology for advanced airborne radar applications \cite{Brennan_Reed_1973}. Although a significant increase in output signal-to-interference-plus-noise-ratio (SINR) compared with traditional approaches can potentially be achieved, the large computational complexity of full-rank STAP, which has order $\mathcal{O}(M^3)$ due to a covariance matrix inversion operation where $M$ is the dimension of the full rank filter, prohibits its application in real-time. Another complicated issue with full rank STAP is that the filter with large number of elements requires a large number of samples to reach its steady-state behaviour while providing slow convergence speed and poor tracking performance \cite{Haykin2002}.
A cost-effective technique in short-data record scenarios and, in particular, for systems containing a large number of elements is reduced-rank signal processing.

In the comprehensive report by Ward \cite{Ward_1994} and the book by Klemm \cite{Klemm_2002_B}, several $ad$ $hoc$ reduced-rank methods were developed. The first statistical reduced-rank method was based on a principal-components (PC) decomposition of the target-free covariance matrix \cite{Haimovich_Bar_1991}. Another class of eigen-decomposition methods was based on the cross-spectral metric (CSM)\cite{Goldstein_Reed_1997_1}. However, the PC and CSM algorithms have the problem of heavy computational load due to the eigen-decomposition. The family of Krylov subspace methods has been investigated thoroughly in the recent years. This class of reduced-rank adaptive filtering algorithms, including the multistage Wiener filter (MSWF) \cite{Goldstein_Reed_Scharf_1998, Goldstein_Reed_Zulch_1999} and auxiliary-vector filters (AVF) \cite{Pados1999,PadosFeb2001,Pados17-20April2007}, projects the observation data onto a lower-dimensional Krylov subspace. Despite the improved convergence and tracking performance achieved with these methods, they are very complex to implement and suffer from numerical problems. The requirement for accurate estimation of the target-free covariance matrix is also an important issue which restricts the statistical reduced-rank STAP algorithms.

In the present paper, we develop a reduced-rank approach to the STAP design utilizing an approach based on adaptive basis function approximation (ABFA). The novel scheme employs the well-known framework of side-lobe canceller (SLC) \cite{Guerci_Goldstein_Reed_2000} structure and consists of selected sets of basis functions that perform dimensionality reduction and an adaptive reduced-rank filter. Compared to traditional reduced-rank techniques, the proposed scheme works on an instantaneous basis, selecting the best suited set of basis functions at each instant to minimize the squared error. Furthermore, we derive stochastic gradient (SG) and recursive least squares (RLS) algorithm for efficiently implementing the proposed ABFA scheme. The results for radar clutter and jamming suppression show a performance significantly better and lower complexity than existing reduced-rank schemes.

This paper is organized as follows. Section \ref{sec:sig_model} introduces the signal model and in Section 3, the proposed reduced-rank STAP scheme is detailed, and the proposed adaptive algorithms and their complexity analysis are also presented in Section 3. Examples of the performance of the proposed reduced-rank STAP are provided in Section 4 and conclusions are given in Section 5.
\section{Signal Model}
\label{sec:sig_model}
The system under consideration is a pulsed Doppler radar residing on an airborne platform. The radar antenna is a uniformly spaced linear array antenna consisting of $N$ elements. Radar returns are collected in a coherent processing interval (CPI), which is referred to as the 3-D radar data-cube shown in Fig.~\ref{fig:radar_data_syst}(a), where $L$ denotes the number of samples collected to cover the range interval. The data is then processed at one range of interest, which corresponds to a slice of the CPI data-cube. This slice is a $J\times N$ matrix which consists of $N\times 1$ spatial snapshots for $J$ pulses at the range of interest. It is convenient to stack the matrix column-wise to form the $M\times 1, M=JN$ vector $\mathbf{r}(i)$, termed a space-time snapshot \cite{Brennan_Reed_1973,Ward_1994}.
Given a space-time snapshot, radar detection is a binary hypothesis problem, where hypothesis $\mathbf{H}_0$ corresponds to target absence and hypothesis $\mathbf{H}_1$ corresponds to target presence. The radar space-time snapshot is then expressed for each of the two hypotheses in the following form,
\begin{equation}
\begin{split}
\mathbf{H}_0&:\mathbf{r}=\boldsymbol{\nu} \\
\mathbf{H}_1&:\mathbf{r}=\alpha\mathbf{s}+\boldsymbol{\nu},
\end{split}
\end{equation}
where $\alpha$ is a complex gain and the vector $\mathbf{s}$, which is the $M\times 1$ normalized space-time steering vector in the space-time look-direction. The vector $\boldsymbol{\nu}$ encompasses any undesired interference or noise component of the data including clutter $\mathbf{c}$, jamming $\mathbf{j}$ and thermal noise $\mathbf{n}$. These three components are assumed to be mutually uncorrelated.
Thus, the $M\times M$ covariance matrix $\mathbf{R}_{\nu}$ of the undesired clutter-plus-jammer-plus-noise component can be modelled as
\begin{equation}
\mathbf{R}_{\nu}=E\{\boldsymbol{\nu}\boldsymbol{\nu}^H\}=\mathbf{R}_c+\mathbf{R}_j+\mathbf{R}_n
\end{equation}
where $H$ represents Hermitian transpose, $\mathbf{R}_c = E\{\mathbf{c}\mathbf{c}^H\}$, $\mathbf{R}_j = E\{\mathbf{j}\mathbf{j}^H\}$ and $\mathbf{R}_n = E\{\mathbf{n}\mathbf{n}^H\}$ denote clutter, jamming and noise covariance matrix respectively. In practice, the interference-plus-noise covariance matrix $\mathbf{R}_{\nu}$ is typically unknown. The common approach is to estimate it from the $secondary$ data set which does not contain the signal of interest ($\mathbf{r}=\boldsymbol{\nu}$). In this context, we can refer the interference-plus-noise covariance matrix $\mathbf{R}_{\nu}$ as $\mathbf{R}$.
The processing chain of a general STAP radar is shown in Fig.~\ref{fig:radar_data_syst}(b). To detect the presence of targets, each range bin is processed by an adaptive 2D beamformer (to achieve maximum output SINR) followed by a hypothesis test to determine the target presence or absence. The optimum full-rank STAP \cite{Brennan_Reed_1973} is given by
\begin{equation}
\boldsymbol{\omega}_{opt}=\gamma\mathbf{R}^{-1}\mathbf{s}
\label{eq:opt_weight}
\end{equation}
where $\gamma$ is a nonzero complex number. In practice, since $\mathbf{R}$ is unknown, the processor substitutes an estimation of $\mathbf{R}$ for $\hat{\mathbf{R}}$ to arrive at the adaptive weight $\hat{\boldsymbol{\omega}}$. It is most common to compute the covariance matrix estimate as $ \hat{\mathbf{R}}$=$\frac{1}{L}\sum_{i=1}^{L}\mathbf{r}(i)\mathbf{r}^H(i)$. This approach is known as sample matrix inversion (SMI).
Assuming Gaussian-distributed interference, an optimum detection statistic follows from the likelihood ratio test and appears as \cite{Melvin2004}
\begin{equation}
\vert d(i)\vert = \vert \boldsymbol{\omega}^H\mathbf{r}(i)\vert {\atopp{\atopp{H_1}{>}}{\atopp{<}{H_0}}}\eta
\label{eq:LRT}
\end{equation}
where $\eta$ is the detection threshold. The performance of (\ref{eq:LRT}) is given by
\begin{equation}
\begin{split}
P_{FA} &= \exp(\frac{-\beta^2}{2}) \\
P_D &= \int_{\beta}^{\infty}\mu\exp\left(\frac{-(\mu^2+\rho^2)}{2}\right)\mathbf{I}_0(\rho\mu)d\mu
\label{eq:PD}
\end{split}
\end{equation}
where $P_{FA}$ is the probability of false alarm, $P_D$ is the probability of detection, $\beta=\eta/\sqrt{\boldsymbol{\omega}^H\mathbf{R}\boldsymbol{\omega}}$ is a normalized detection threshold, $\mathbf{I}_0(\cdot)$ is the modified zero-order Bessel function of the first kind, and $\rho$ equals the square-root of the peak output SINR. According to (\ref{eq:PD}), $P_D$ depends on both output SINR and the value of $P_{FA}$ in a white noise detection scenario. If the value of $P_{FA}$ is specified, $P_D$ is a monotonic function of $\rho$. Thus, maximizing SINR likewise maximizes $P_D$ for a fixed value of $P_{FA}$.
\section{Proposed Algorithm}
\label{sec:proposed_algorithm}
In this section, a novel adaptive reduced-rank scheme based on adaptive basis function approximation (ABFA) is proposed. Then, we derive the SG and RLS algorithms for efficiently implementing the proposed scheme. Finally, we compare the computational complexity in terms of multiplications of the proposed scheme and the existing algorithms.
\subsection{Algorithm Description}
The novel scheme employs the well-known framework of SLC structure and consists of selected sets of basis functions that perform dimensionality reduction and an adaptive reduced-rank filter. As shown in Fig.~\ref{fig:ABFA_RR}, we begin by using a desired target signal $\mathbf{s}$ (steering vector) as a first-stage matched filter to form a main-beam response $d_0(i)=\mathbf{s}^H\mathbf{r}(i)$. In our design, there are $B$ sets of basis functions $\{\mathbf{T}_b\in\mathcal{C}^{M\times D}\vert b=1,...,B\}$ pre-stored as candidate projection matrices, where $D$ is the length of reduced-rank filter. Based on Fig.~\ref{fig:ABFA_RR}, we have a set of values $\{y_b(i)\}$ at the output of SLC which can be written as a function of $\bar{\boldsymbol{\omega}}(i)$ and the instantaneous projection matrix $\mathbf{T}_b(i)$ as
\begin{equation}
y_b(i)=\bar{\boldsymbol{\omega}}^H(i)\bar{\mathbf{r}}(i)=\bar{\boldsymbol{\omega}}^H(i)\mathbf{T}_b^H(i)\mathbf{r}^{\prime}(i)
\end{equation}
where $\bar{\boldsymbol{\omega}}(i)$ denotes the reduced-rank filter weight vector, $\mathbf{T}_b(i)$ denotes the $b$th set of basis functions and $\mathbf{r}^{\prime}(i) = \mathbf{B}\mathbf{r}(i)$ is the received data projected in the null space of $\mathbf{s}$, $\mathbf{B}=\mathbf{null}(\mathbf{s})$ or $\mathbf{I}-\frac{\mathbf{s}\mathbf{s}^H}{\Vert\mathbf{s}\Vert^2}$ is the block matrix. At the $i$th time instant, we select one matrix from these $B$ sets of basis functions as the instantaneous projection matrix by using the decision rule
\begin{equation}
b_{opt}=\arg\min_{b\in \{1,...,B\}}\vert e_b(i)\vert^2
\end{equation}
where $e_b(i)=d_0(i)-y_b(i)$.
The pre-stored matrices are constructed by basis functions in the following way
\begin{equation}
\mathbf{T}_b=[\phi_{b,1} \quad \phi_{b,2} \quad...\quad \phi_{b,D}]
\end{equation}
where $\phi_{b,d}\in \mathcal{C}^{M\times 1}$ denotes the $d$th basis function of the $b$th set, $d=1,...,D$ and $b=1,...,B$ which is composed of a single "one" and $(M-1)$ zeros, according to the following
\begin{equation}
\phi_{b,d}=[\underbrace{0 \quad ... \quad 0}_{z_{b,d}}\quad 1 \quad \underbrace{0 \quad ... \quad 0}_{M-z_{b,d}-1}]
\end{equation}
where $z_{b,d}$ is the number of zeros before the only element equal to one. We set the value of $z_{b,d}$ in a deterministic way which can be expressed as
\begin{equation}
z_{b,d}=\frac{M}{D}\times (d-1)+(b-1).
\end{equation}
It should be remarked that other designs have been investigated and this structure has been adopted due to an excellent trade-off between performance and complexity.
\subsection{Adaptation Implementations}
Here, we describe the SG and RLS algorithms that adjust the parameters of the reduced-rank filter based
on the MSE and LS cost function.
\subsubsection{The SG Algorithm}
We present a low-complexity SG adaptive reduced-rank algorithm for efficiently implementing the proposed method. The reduced-rank MSE cost function is given by
\begin{equation}
\mathcal{J}_{\mathbf{MSE}}(\bar{\boldsymbol{\omega}}(i))=\mathbf{E}\left\{\vert d_0(i)-\bar{\boldsymbol{\omega}}^H(i)\mathbf{T}^H(i)\mathbf{r}^{\prime}(i)\vert^2\right\}.
\label{eq:cost_function}
\end{equation}
By computing the instantaneous gradient terms of (\ref{eq:cost_function}) with respect to $\bar{\boldsymbol{\omega}}(i)$ and introducing the positive step size $\mu$, the resulting ABFA-SG algorithm is thus given by
\begin{eqnarray}
y_b(i) &=& \bar{\boldsymbol{\omega}}^H(i)\mathbf{T}_b^H(i)\mathbf{r}^{\prime}(i), \\
b_{opt}&=& \arg\min_{b\in \{1,...,B\}}\vert d_0(i)-y_b(i)\vert^2, \\
\bar{\mathbf{r}}(i)&=&\mathbf{T}_{b_{opt}}^H\mathbf{r}^{\prime}(i), \\
e_o(i)&=& d_0(i)-\bar{\boldsymbol{\omega}}^H(i)\bar{\mathbf{r}}(i), \\
\bar{\boldsymbol{\omega}}(i+1) &=& \bar{\boldsymbol{\omega}}(i) +\mu\bar{\mathbf{r}}(i)e_o^*(i).
\end{eqnarray}
\subsubsection{The RLS Algorithm}
Here, we develop the RLS-version of the proposed ABFA scheme. To this end, let us first consider the LS design of the proposed scheme as
\begin{equation}
\mathcal{J}_{\mathbf{LS}}(\bar{\boldsymbol{\omega}}(i))=\sum_{j=1}^{i}\lambda^{i-j}\vert d_0(j)-\bar{\boldsymbol{\omega}}^H(i)\mathbf{T}^H(i)\mathbf{r}^{\prime}(j)\vert^2
\label{eq:LS_cost}
\end{equation}
where $\lambda$ is the forgetting factor. By computing the gradient of (\ref{eq:LS_cost}) with respect to $\bar{\boldsymbol{\omega}}(i)$ equal to zero, we obtain
\begin{equation}
\bar{\boldsymbol{\omega}}(i) = \bar{\mathbf{R}}^{-1}\bar{\mathbf{p}},
\end{equation}
where $\bar{\mathbf{R}} = \sum_{j=1}^{i}\lambda^{i-j}\bar{\mathbf{r}}(j)\bar{\mathbf{r}}(j)^H$ and $\bar{\mathbf{p}} = \sum_{j=1}^{i}\lambda^{i-j}d_0^*(j)\bar{\mathbf{r}}(j)$ denote the time averaged correlation matrix and cross-correlation vector respectively. Thus the optimal weight can be calculated recursively using the RLS algorithm, which is summarized as follows
\begin{eqnarray}
\mathbf{P}(0)&=&\delta^{-1}\mathbf{I}, \\
\mathbf{K}(i)&=&\frac{\mathbf{P}(i-1)\bar{\mathbf{r}}(i)}{\lambda+\bar{\mathbf{r}}^H(i)\mathbf{P}(i-1)\bar{\mathbf{r}}(i)}, \\
\mathbf{P}(i)&=&\lambda^{-1}\mathbf{P}(i-1)+\lambda^{-1}\mathbf{K}(i)\bar{\mathbf{r}}^H(i)\mathbf{P}(i-1), \\
\bar{\boldsymbol{\omega}}(i) &=& \bar{\boldsymbol{\omega}}(i-1) + \mathbf{K}(i)e_o^*(i)
\end{eqnarray}
where $\delta$ is a small constant and $\mathbf{I}$ is the identity matrix.
\begin{table}
\renewcommand{\baselinestretch}{0.75}
\centering
\caption{Computational complexity of algorithms.}
\begin{tabular}{l c}
\hline\hline
Algorithm & Multiplications \\[0.5ex]
\hline
Full-Rank-SG&$2M+1$ \\[0.5ex]
Full-Rank-RLS&$4M^2+4M$\\[0.5ex]
\multirow{2}{*}{MSWF-SG}&$(D+1)M^2+D$ \\[0.5ex]
&$+(3D+2)M$\\[0.5ex]
\multirow{2}{*}{MSWF-RLS}&$(D+1)M^2+3DM$ \\[0.5ex]
&$+2M+4(D^2+D)$\\[0.5ex]
\multirow{2}{*}{AVF} &$D(4M^2+4M+1)$ \\[0.5ex]
& $+4M+2$\\[0.5ex]
\multirow{2}{*}{ABFA-SG} & $(B+3)D+2$\\[0.5ex]
& $M^2+M$ \\[0.5ex]
\multirow{2}{*}{ABFA-RLS} & $4D^2+(B+3)D+B$ \\[0.5ex]
& $M^2+M$ \\[0.5ex]
\hline
\end{tabular}
\renewcommand{\baselinestretch}{1.0}
\label{tab:complexity}
\end{table}
\subsection{Complexity Analysis}
Here, we detail the computational complexity in terms of multiplications of the proposed schemes with SG and RLS
and other existing algorithms, namely the full-rank with SG
and RLS, the MSWF with SG and RLS and the AVF,
as shown in Table \ref{tab:complexity}, where $D$ is the length of the reduced-rank filter and $B$ is the number of sets of pre-stored basis functions in the proposed scheme. The proposed scheme with both SG and RLS algorithms is
much simpler than the full-rank with RLS, the MSWF and the
AVF and slightly more complex than the Full-rank with SG
(for $D \ll M$). To compare the complexity of our proposed algorithms with other reduced-rank algorithms, let us take a simple example, say $M=64$, $D=4$ and $B=16$, the number of multiplications for MSWF with SG is 21380, for MSWF with RLS is 21456 and for AVF is 66822 respectively, while our proposed algorithm has significant lower number of multiplications 4233 for SG and 4316 for RLS respectively.

\section{Simulations}
\begin{figure}
\centering
\includegraphics[width=3.4in]{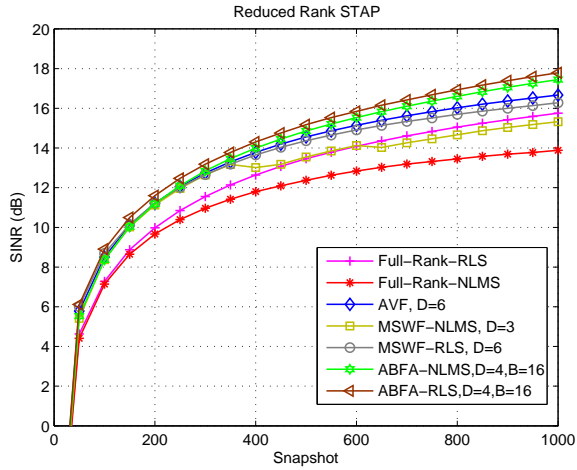}
\caption{SINR performance against snapshot with $M$ = 64, SNR = 0 dB, $\lambda$ = 0.9998, $\mu$ = 0.005.}
\label{fig:SINRvsSS}
\end{figure}

\begin{figure}
\centering
\includegraphics[width=3.4in]{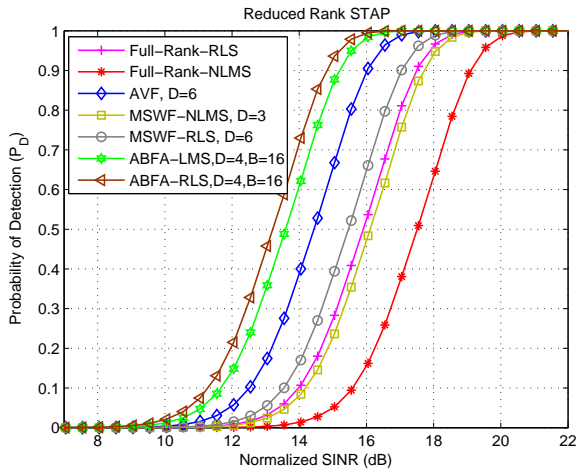}
\caption{Probability of detection performance vs normalized SINR with $M$ = 64, $\lambda$ = 0.9998, $\mu$ = 0.005, $L$ = 1000 snapshots.}
\label{fig:SINRvsPD}
\end{figure}

\begin{table}
\centering
\caption{Radar System Parameters}
\begin{tabular}{l c}
\hline\hline
Parameter & Value\\ [0.5ex]
\hline
Carrier frequency ($f_c$) & 450 MHz \\
Transmit pattern & Uniform \\
PRF ($f_r$) & 300 Hz \\
Platform velocity ($v$) & 50 m/s \\
Platform height ($h$) & 9000 m \\
Clutter-to-Noise ratio (CNR) & 40 dB \\
Jammer-to-Noise ratio (JNR) & 30 dB \\
Elements of sensors ($N$) & 8\\
Number of Pulses ($M$) & 8 \\[1ex]
\hline
\end{tabular}
\label{tab:Radar_system_parameters}
\end{table}
In this section we assess the proposed ABFA STAP algorithm in an airborne radar application. The parameters of the radar platform are shown in the Table~\ref{tab:Radar_system_parameters}. For all simulations, we assume the presence of a mixture of two broadband jammers at $-45^\circ$ and $60^\circ$ with Jammer-to-Noise-ratio (JNR) equal to 30 dB. The Clutter-to-Noise-ratio (CNR) is fixed at 40 dB. $L = 1000$ snapshots are simulated and all presented results are averages over 100 independent Monte-Carlo runs.
We compare the proposed scheme with the full-rank filter, the MSWF and the AVF techniques for the design of linear receivers, where the reduced-rank filter $\bar{\boldsymbol{\omega}}(i)$ with $D$ coefficients provides an estimate to determine whether the target is present or not. The signal-to-noise-ratio is set at 0 dB. We evaluate the SINR performance of our proposed algorithm against the number of snapshots $L$ by comparing with all other schemes, which is shown in Fig.~\ref{fig:SINRvsSS}. The curves show an excellent performance for the proposed scheme, which converges much faster and has significantly better performance than the full-rank filter, the MSWF and the AVF schemes at much lower complexity. In the second experiment, we calculate $P_D$ based on (\ref{eq:PD}) by setting the false alarm at $P_{FA}=10^{-10}$. Fig.~\ref{fig:SINRvsPD} presents $P_D$ versus normalized SINR results for all schemes. The figure illustrates that the best detection performance is provided by the proposed scheme using RLS and SG adaptive algorithms, followed by the AVF and the MSWF, and finally the full-rank filters.
\section{Conclusions}
In this paper, we developed a reduced-rank approach to the STAP design utilizing an approach based on adaptive basis function approximation (ABFA). The novel scheme employed the well-known framework of the SLC structure and consists of selected sets of basis functions that perform dimensionality reduction and an adaptive reduced-rank filter. Compared to traditional reduced-rank techniques, the proposed scheme worked on an instantaneous basis, selecting the best suited set of basis functions at each instant to minimize the squared error. Furthermore, we derived the SG and RLS algorithm for efficiently implementing the proposed ABFA scheme. The results for radar clutter and jamming suppression showed a performance significantly better and lower complexity than existing reduced-rank schemes.

\begin{thebibliography}{10}

\footnotesize
\bibitem{Brennan_Reed_1973}
L.~E. Brennan and L.~S. Reed,
\newblock ``Theory of adaptive radar,''
\newblock {\em IEEE Trans. Aero. and Elect. Syst.}, vol.
  AES-9, no. 2, pp. 237--252, 1973.

\bibitem{Haykin2002}
S.~Haykin,
\newblock {\em Adaptive Filter Theory},
\newblock NJ: Prentice-Hall, 4th ed., 2002.

\bibitem{Ward_1994}
J.~Ward,
\newblock ``Space-time adaptive processing for airborne radar,,''
\newblock {\em Technical report 1015, MIT Lincoln laboratory, Lexington,MA},
  Dec. 1994.

\bibitem{Klemm_2002_B}
R.~Klemm,
\newblock {\em Principle of Space-Time Adaptive Processing},
\newblock IEE Press, Bodmin, UK, 2002.

\bibitem{Haimovich_Bar_1991}
A.~M. Haimovich and Y.~Bar-Ness,
\newblock ``An eigenanalysis interference canceler,''
\newblock {\em  IEEE Trans. Signal Process.}, vol. 39, no. 1, pp.
  76--84, 1991.

\bibitem{Goldstein_Reed_1997_1}
J.~S. Goldstein and I.~S. Reed,
\newblock ``Reduced-rank adaptive filtering,''
\newblock {\em IEEE Trans. Signal Process.}, vol. 45, no. 2, pp.
  492--496, 1997.

\bibitem{Goldstein_Reed_Scharf_1998}
J.~S. Goldstein, I.~S. Reed, and L.~L. Scharf,
\newblock ``A multistage representation of the wiener filter based on
  orthogonal projections,''
\newblock {\em  IEEE Trans. Inf. Theory}, vol. 44, no. 7, pp.
  2943--2959, 1998.

\bibitem{Goldstein_Reed_Zulch_1999}
J.~S. Goldstein, I.~S. Reed, and P.~A. Zulch,
\newblock ``Multistage partially adaptive STAP CFAR detection algorithm,''
\newblock {\em  IEEE Trans. Aero. and Elect. Syst.}, vol.
  35, no. 2, pp. 645--661, 1999.

\bibitem{Pados1999}
D.A. Pados and S.N. Batalama,
\newblock ``Joint space-time auxiliary-vector filtering for DS-CDMA systems
  with antenna arrays,''
\newblock {\em IEEE Trans. Commun.}, vol. 47, no. 9, pp.
  1406--1415, 1999.

\bibitem{PadosFeb2001}
D.A. Pados and G.N. Karystinos,
\newblock ``An iterative algorithm for the computation of the MVDR filter,''
\newblock {\em  IEEE Trans. Signal Process.}, vol. 49, no. 2, pp.
  290--300, Feb 2001.

\bibitem{Pados17-20April2007}
D.A. Pados, G.N. Karystinos, S.N. Batalama, and J.D. Matyjas,
\newblock ``Short-data-record adaptive detection,''
\newblock {\em Radar Conference, 2007 IEEE}, pp. 357--361, 17-20 April 2007.

\bibitem{Guerci_Goldstein_Reed_2000}
J.~R. Guerci, J.~S. Goldstein, and I.~S. Reed,
\newblock ``Optimal and adaptive reduced-rank STAP,''
\newblock {\em  IEEE Trans. Aero. and Elect. Syst.}, vol.
  36, no. 2, pp. 647--663, 2000.

\bibitem{Melvin2004}
W.L. Melvin,
\newblock ``A STAP overview,''
\newblock {\em IEEE Aero. and Elect. Syst. Mag.}, vol. 19, no. 1,
  pp. 19--35, 2004.

\bibitem{delamaresp}
R. C. de Lamare and R. Sampaio-Neto, ``Adaptive Reduced-Rank MMSE
Filtering with Interpolated FIR Filters and Adaptive Interpolators",
\textit{IEEE Sig. Proc. Letters}, vol. 12, no. 3, March, 2005.

\bibitem{delamarecl}
R. C. de Lamare and Raimundo Sampaio-Neto, ``Reduced-rank
Interference Suppression for DS-CDMA based on Interpolated FIR
Filters", \textit{IEEE Communications Letters}, vol. 9, no. 3, March
2005.

\bibitem{delamaretvt}
R. C. de Lamare and R. Sampaio-Neto, ``Adaptive Interference
Suppression for DS-CDMA Systems based on Interpolated FIR Filters
with Adaptive Interpolators in Multipath Channels", \textit{IEEE
Transactions on Vehicular Technology}, Vol. 56, no. 6, September
2007.

\bibitem{delamarespl07}
R. C. de Lamare and R. Sampaio-Neto, ``Reduced-Rank Adaptive
Filtering Based on Joint Iterative Optimization of Adaptive
Filters", \textit{IEEE Sig. Proc. Letters}, Vol. 14, no. 12,
December 2007.

\bibitem{delamaretvt10}
R. C. de Lamare and R. Sampaio-Neto, ``Reduced-Rank Space-Time
Adaptive Interference Suppression With Joint Iterative Least Squares
Algorithms for Spread-Spectrum Systems," \textit{IEEE Transactions
on Vehicular Technology}, vol.59, no.3, March 2010, pp.1217-1228.

\bibitem{jidf_icassp}
R. C. de Lamare and R. Sampaio-Neto, ``Adaptive Reduced-Rank MMSE
Parameter Estimation based on an Adaptive Diversity Combined
Decimation and Interpolation Scheme," \textit{Proc. IEEE
International Conference on Acoustics, Speech and Signal
Processing}, April 15-20, 2007, vol. 3, pp. III-1317-III-1320.

\bibitem{jidf}
R. C. de Lamare and R. Sampaio-Neto, ``Adaptive Reduced-Rank
Processing Based on Joint and Iterative Interpolation, Decimation,
and Filtering," \textit{IEEE Transactions on Signal Processing},
vol. 57,  no. 7,  July 2009, pp. 2503 - 2514.

\bibitem{barc}
R.C. de Lamare, R. Sampaio-Neto, M. Haardt, "Blind Adaptive
Constrained Constant-Modulus Reduced-Rank Interference Suppression
Algorithms Based on Interpolation and Switched Decimation,"
\textit{IEEE Transactions on Signal Processing}, vol.59, no.2,
pp.681-695, Feb. 2011.

\bibitem{delamareelb}
R. C. de Lamare, ``Adaptive Reduced-Rank LCMV Beamforming Algorithms
Based on Joint Iterative Optimisation of Filters",
\textit{Electronics Letters}, vol. 44, no. 9, 2008.



\bibitem{jidf}
R. C. de Lamare and R. Sampaio-Neto, ``Adaptive Reduced-Rank
Processing Based on Joint and Iterative Interpolation, Decimation
and Filtering", \textit{IEEE Transactions on Signal Processing},
vol. 57, no. 7, July 2009, pp. 2503 - 2514.

\bibitem{delamare10}
R. C. de Lamare, L. Wang, and R. Fa, ``Adaptive reduced-rank LCMV
beamforming algorithms based on joint iterative optimization of
filters: Design and analysis," Signal Processing, vol. 90, no. 2,
pp. 640-652, Feb. 2010.

\bibitem{fa10}
R. Fa, R. C. de Lamare, and L. Wang, ``Reduced-Rank STAP Schemes for
Airborne Radar Based on Switched Joint Interpolation, Decimation and
Filtering Algorithm," \textit{IEEE Transactions on Signal
Processing}, vol.58, no.8, Aug. 2010, pp.4182-4194.

\end{thebibliography}

\end{document}